 \documentclass[smallabstract,smallcaptions]{dccpaper}

\usepackage{epsfig}
\usepackage{citesort}
\usepackage{amsmath}
\usepackage{amssymb}
\usepackage{color}
\usepackage{url}
\usepackage{algorithm}
\usepackage{algpseudocode}
\usepackage{fancyhdr}
\usepackage[braket, qm]{qcircuit}
\usepackage{graphicx}
\usepackage{tikz}
\usepackage{dirtytalk}
\usetikzlibrary{positioning, fit, decorations.pathreplacing}

\newlength{\figurewidth}
\newlength{\smallfigurewidth}

\begin{document}

\title
{\large
\textbf{Lossy Compression for Schrödinger-style Quantum Simulations}
}

\author{%
Noah Huffman$^{\ast\ddag}$, Dmitri Pavlichin, and Tsachy Weissman$^{\ast}$\\[0.5em]
{\small
  \begin{minipage}{\linewidth}
    \begin{center}
      \begin{tabular}{cc}
        $^{\ast}$Stanford University &  $^{\ddag}$Corresponding Author\\
        Palo Alto, CA, 94306, USA &  \url{nhuffman@stanford.edu}\\
      \end{tabular}
    \end{center}
  \end{minipage}
}
}

\maketitle
\thispagestyle{empty}

\begin{abstract}
Simulating quantum circuits on classical hardware is a powerful and necessary tool for developing and testing quantum algorithms and hardware as well as evaluating claims of quantum supremacy in the Noisy Intermediate-Scale Quantum (NISQ) regime. Schrödinger-style simulations are limited by the exponential growth of the number of state amplitudes which need to be stored. In this work, we apply scalar and vector quantization to Schrödinger-style quantum circuit simulations as lossy compression schemes to reduce the number of bits needed to simulate quantum circuits. Using quantization, we can maintain simulation fidelities $>0.99$ when simulating the Quantum Fourier Transform, while using only 7 significand bits in a floating-point number to characterize the real and imaginary components of each amplitude. Furthermore, using vector quantization, we propose a method to bound the number of bits/amplitude needed to store state vectors in a simulation of a circuit that achieves a desired fidelity, and show that for a 6 qubit simulation of the Quantum Fourier Transform, 15 bits/amplitude is sufficient to maintain fidelity $>0.9$ at $10^4$ depth.
\end{abstract}

\Section{Introduction}
\par Quantum computers are state-of-the-art machines, which are known to solve problems that are hard for classical computers to solve \cite{shor1999algorithms}. While quantum computers that can solve meaningful problems exist, implementation of quantum computers is currently limited by practical hardware challenges \cite{preskill2012quantum,harrow2017quantum,boixo2018characterizing,arute2019quantum}.  Due to these complications, quantum computers of the immediate future will be notably small and noisy, leading to what many refer to as the Noisy Intermediate-Scale Quantum (NISQ) regime \cite{preskill2018quantum}.  During this period, simulations of quantum computers and systems on classical hardware are of great interest because they allow us to explore problems that our current quantum computers are too small and noisy to explore, and because they raise the bar for verification of Quantum Supremacy claims \cite{jones2019quest}.

\SubSection{Quantum Computation Fundamentals}

A qubit can be represented as a 2-dimentional vector with complex coefficients, $\left\{\psi_1, \psi_2\right\} \in \mathbb{C}$, in other words:

\begin{equation}
\resizebox{0.5\linewidth}{!}{
    $\begin{pmatrix}
        \psi_1 \\
        \psi_2
    \end{pmatrix}
    = \psi_1\begin{pmatrix}
        1 \\
        0
    \end{pmatrix}
    + \psi_2\begin{pmatrix}
        0 \\
        1
    \end{pmatrix}
    = \psi_1|0\rangle + \psi_2|1\rangle = |\psi\rangle$
}
\end{equation}

Here, the basis vectors, $|0\rangle$ and 1$\rangle$, serve as the computational basis. Thus, the vector describing the qubit can be seen as a superposition of these two basis vectors with complex coefficients and the additional normalization constraint:

\begin{equation}
    \left|\psi_1\right|^2+\left|\psi_2\right|^2=1
\end{equation}

More generally, an $n$-qubit system can be described as a superposition of $N = 2^n$ basis states, each with its own complex amplitude and subject to a similar normalization constraint:

\begin{equation}
    |\psi\rangle=\psi_1|000 \ldots 00\rangle+\psi_2|000 \ldots 01\rangle+\cdots+\psi_{N}|111 \ldots 111\rangle
\end{equation}


\begin{equation}
    \forall \psi_i \in \mathbb{C} \quad \& \quad \sum_{i=1}^N\left|\psi_i\right|^2=1
\end{equation}

To perform computations in the Schrödinger paradigm, these quantum states are acted upon by $2^n \times 2^n$ unitary matrices (i.e. \say{gates}), which are the tensor products of smaller gates that affect smaller subsets of the total qubit arrangement. For example, applying a single-qubit gate, $U$, to the $k^{\text {th }}$ qubit in a quantum circuit is represented by the $2^n \times 2^n$ unitary transformation, $A=I^{\otimes n-k-1} \otimes U \otimes I^{\otimes k}$.

\par Our lossy compression schemes strive to preserve the quantum fidelity between the original vector and our lossy reconstruction. Given two density matrices, $\rho$ and $\sigma$, the fidelity is defined as $f(\rho, \sigma)=(\operatorname{tr} \sqrt{\sqrt{\rho} \sigma \sqrt{\rho}})^2$. For the specific case where both quantum states are pure (i.e. $\rho=\left|\psi_\rho\right\rangle\left\langle\psi_\rho\right|$ and $\left.\sigma=\left|\psi_\sigma\right\rangle\left\langle\psi_\sigma\right|\right)$,  the fidelity reduces to $f(\rho, \sigma)=\left|\left\langle\psi_\rho \mid \psi_\sigma\right\rangle\right|^2$.

\SubSection{Related Work}

\par Several methods exist to simulate quantum circuits on classical hardware. The most fundamental of these is Schrödinger-style simulation, which maintains the full $2^n$ quantum state vector in memory and updates the state vector sequentially by applying the unitary transformations defined by gates which act upon the qubits (see \textit{Quantum Computation Fundamentals} above). Schrödinger simulations can further be segregated into array methods and decision diagram methods \cite{burgholzer2021hybrid,tsai2016bit}. In general, Schrödinger methods scale both exponentially in time and space with the number of qubits, as both storing and operating on the full $2^n$ quantum state vector is proportional to its size \cite{viamontes2008gate,frank2009space}. However, since the state vector only needs to be updated $d$ times, where $d$ is the number of gates in the circuit (i.e. circuit depth), Schrödinger-style simulation only scales linearly with circuit depth. Currently, state-of-the-art Schrödinger simulations can calculate circuits with $n\approx50$  \cite{zulehner2019advanced,wu2019compression,betelu2020limits}.

\par Alternatively, Feynman-style simulations rely on summation over integral paths \cite{wu2019compression,burgholzer2021hybrid}. Effectively, Feynman path simulations work by considering each gate which connects two or more qubits as a decision point from which the simulation branches. Feynman simulations compute an amplitude by summing the contributions of every possible “path” through the decision tree to compute the final result \cite{wu2019compression,burgholzer2021hybrid}. These simulations are only polynomial in space \cite{aaronson2017complexity}, but because the number of paths scales exponentially with the number of decision points these simulations require $\Theta(2^{dn})$ time \cite{wu2019compression,burgholzer2021hybrid}, which means that they are impractical for deep circuit simulation.

Finally, tensor network approaches represent quantum circuits' connectivity in a tensor network. While these methods are popular, they only work well if the connectivity of the gates is reduced to a grid, or the simulation handles states close to linear combinations of product states \cite{betelu2020limits}. Furthermore, the time and space costs for contracting these tensor networks scales exponentially with the treewidth of the underlying graphs, and therefore are impractical to simulate deep quantum circuits \cite{wu2019compression,betelu2020limits}. Furthermore, tensor-network based simulations often only simulate a small number of amplitudes instead of the entire state vector \cite{burgholzer2021hybrid}.

\par Of these methods, Schrödinger-style simulation is the most appealing setting for compression, as it scales well with circuit depth, but is held back by memory requirements. Work has been done to explore the use of lossy compression on classical simulations to improve the space requirements of said simulations \cite{wu2018amplitude,wu2019compression}. These initial results found that traditional lossy compressors such as SZ and ZFP underperform on this task due to the lack of patterns or “spikiness” in the quantum data. These same results also show that bit truncation may serve as a promising method for lossy data compression \cite{wu2019compression}. Moreover, benefits of low-precision simulation for quantum circuits have been previously explored for non-native algebraic representations of complex values \cite{betelu2020limits}, but such representations require conversion to double precision to perform all arithmetic operations (i.e. gate applications). We expand upon this literature by performing traditional floating-point Schrödinger-style simulation compressed with scalar and vector quantized amplitude values.

\par In this work, we consider the computation of all complex amplitudes of the quantum state resulting from the execution of the circuit as well as intermediate steps.  Furthermore, we are concerned with techniques to simulate general quantum circuits, so we eschew specialized simulation techniques which assume reduced gate sets \cite{gottesman1998heisenberg,wu2019compression}.

\Section{Lossy Compression}

\SubSection{Scalar Quantization}

\par We begin by noting that matrix multiplication can be conjugated into multiplication by constituent real and imaginary components

\begin{equation}
\label{scalar_mult}
U \vec{\psi}=(R+i J)(\vec{a}+i \vec{b})=(R \vec{a}-J \vec{b})+i(J \vec{a}+R \vec{b})
\end{equation}

\begin{figure}
\centering
\resizebox{0.6\linewidth}{!}{%
    \begin{tabular}{c}
    \begin{tikzpicture}[scale=0.5]
        
        \tikzstyle{blueblock} = [draw, fill=blue!20, text centered, minimum height=1em, text width=1em]
        \tikzstyle{greenblock} = [draw, fill=green!20, text centered, minimum height=1em, text width=1em]
        \tikzstyle{redblock} = [draw, fill=red!20, text centered, minimum height=1em, text width=1em]
        \tikzstyle{brace} = [decorate, decoration={brace, amplitude=5pt}]
        
        \node [blueblock] (block1) at (0,0) {};
        \draw [brace] (block1.north west) -- (block1.north east) node[midway, above=5pt] {Sign};
        
        \foreach \x in {1,...,5} {
          \pgfmathtruncatemacro{\nextx}{\x + 1}
          \node [greenblock, right=0em of block\x] (block\nextx) {};
        }
        
        \foreach \x in {7,...,16} {
          \pgfmathtruncatemacro{\prevx}{\x - 1}
          \node [redblock, right=0em of block\prevx] (block\x) {};
        }
        
        \draw [brace] (block2.north west) -- (block6.north east) node[midway, above=5pt] {Exponent (5 bits)};
        
        \draw [brace] (block7.north west) -- (block16.north east) node[midway, above=5pt] {Significand (10 bits)};
        
        \end{tikzpicture} \\
        {\small (a) IEEE float16 half-precision format \cite{IEEEprecision}} \\
    \begin{tikzpicture}[scale=0.5]
        
        \tikzstyle{blueblock} = [draw, fill=blue!20, text centered, minimum height=1em, text width=1em]
        \tikzstyle{greenblock} = [draw, fill=green!20, text centered, minimum height=1em, text width=1em]
        \tikzstyle{redblock} = [draw, fill=red!20, text centered, minimum height=1em, text width=1em]
        \tikzstyle{brace} = [decorate, decoration={brace, amplitude=5pt}]
        
        \node [blueblock] (block1) at (0,0) {};
        \draw [brace] (block1.north west) -- (block1.north east) node[midway, above=5pt] {Sign};
        
        \foreach \x in {1,...,8} {
          \pgfmathtruncatemacro{\nextx}{\x + 1}
          \node [greenblock, right=0em of block\x] (block\nextx) {};
        }
        
        \foreach \x in {10,...,16} {
          \pgfmathtruncatemacro{\prevx}{\x - 1}
          \node [redblock, right=0em of block\prevx] (block\x) {};
        }
        
        \draw [brace] (block2.north west) -- (block9.north east) node[midway, above=5pt] {Exponent (8 bits)};
        
        \draw [brace] (block10.north west) -- (block16.north east) node[midway, above=5pt] {Significand (7 bits)};
        
        \end{tikzpicture} \\
    {\small (b) bfloat16 format}\\
        \begin{tikzpicture}[scale=0.5]
        
        \tikzstyle{blueblock} = [draw, fill=blue!20, text centered, minimum height=1em, text width=1em]
        \tikzstyle{greenblock} = [draw, fill=green!20, text centered, minimum height=1em, text width=1em]
        \tikzstyle{redblock} = [draw, fill=red!20, text centered, minimum height=1em, text width=1em]
        \tikzstyle{blackblock} = [draw, fill=gray!20, text centered, minimum height=1em, text width=1em, text=white]
        \tikzstyle{brace} = [decorate, decoration={brace, amplitude=5pt}]
        
        \node [blueblock] (block1) at (0,0) {};
        \draw [brace] (block1.north west) -- (block1.north east) node[midway, above=5pt] {Sign};
        
        \foreach \x in {1,...,5} {
          \pgfmathtruncatemacro{\nextx}{\x + 1}
          \node [greenblock, right=0em of block\x] (block\nextx) {};
        }
    
        \foreach \x in {7,...,14} {
          \pgfmathtruncatemacro{\prevx}{\x - 1}
          \node [redblock, right=0em of block\prevx] (block\x) {};
        }
        
        \foreach \x in {15,16} {
          \pgfmathtruncatemacro{\prevx}{\x - 1}
          \node [blackblock, right=0em of block\prevx] (block\x) {};
        }

        \draw [brace] (block2.north west) -- (block6.north east) node[midway, above=5pt] {Exponent (5 bits)};
    
        \draw [brace] (block7.north west) -- (block14.north east) node[midway, above=5pt] {Significand (8 bits)};
    
        \draw [brace] (block15.north west) -- (block16.north east) node[midway, above=5pt] {not used};

        \end{tikzpicture} \\
        {\small (c) \say{float8} custom precision}\\
    \end{tabular}
}
\caption{16 bit precisions utilized. (a) standard float16 format (b) bfloat16 dedicates more bits to the exponent than float16, and can take on the same range of values as float32, making it a popular tool in machine learning because conversion to and from float32 is simple \cite{Kalamkar2019}. (c) Our custom ``floatk'' precision forces the use of only $k$ significand bits in a 16 bit float. The exponent is 5 bits, just like float16.}
\label{fig:precision}
\end{figure}
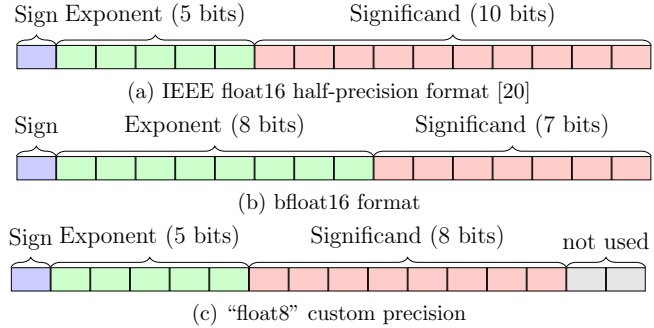

We quantize each float in $R$, $J$, $\vec{a}$, and $\vec{b}$ separately, then perform the multiplications in equation \ref{scalar_mult} for each gate application in the circuit (Algorithm \ref{alg:Gate}). For $n$ qubits, $R$ and $J$ each have $2^{2n}$ floating point values, while  $\vec{a}$ and $\vec{b}$ each have $2^n$ floating point values. We compress the real and imaginary components of each complex number to the following precisions: double precision (float64, 128 bits/complex number), single precision (float32, 64 bits/complex number), half precision (float16, 32 bits/complex number), and bfloat16 (Figure \ref{fig:precision}), a precision popular in machine learning \cite{Kalamkar2019,Google2019bfloat16}. Additionally, we explored extreme bit truncation by rounding the significand of float16 objects to $k$ bits. These significand-truncated precisions we refer to as \say{floatk} where $k$ is the number of usable bits in the significand (Figure \ref{fig:precision}).

\begin{algorithm}
    \caption{Gate Apply}\label{alg:Gate}
    \resizebox{0.76\linewidth}{!}{%
        \begin{minipage}{\linewidth}
            \begin{algorithmic}
                \Require Gate ($U$), state vector ($\psi$), and desired precision
                \State $R \leftarrow real(U)$
                \State $J \leftarrow imag(U)$
                \State $a \leftarrow real(\psi)$
                \State $b \leftarrow imag(\psi)$
                \For{object in $\left[ R, J, a, b \right]$}
                    \For{float in object}
                        \State $float \leftarrow cast(float,\text{desired precision})$
                    \EndFor
                \EndFor \\
                \Return $\psi_{out} = (R \cdot a-J \cdot b)+i(J \cdot a+R \cdot b)$
            \end{algorithmic}
        \end{minipage}%
    }
\end{algorithm}

\SubSection{Vector Quantization}

\par We next utilize vector quantization to lossly represent quantum data. Because quantum state vectors, $\ket{\psi}$, consist of $2^n$ complex amplitudes, quantizing these amplitudes in the complex plane is natural and powerful means of representing the data. We consider the setting where the circuit meant to be compressed is one which is a frequently-used subroutine of many quantum algorithms. We employ a \say{two-pass} vector quantization solution (Algorithm \ref{alg:2pass}) where the circuit is first run once without compression, and then those complex amplitude values are used to create codewords for future runs of the same circuit. By using this \say{two-pass} method, we avoid the need to run k-means multiple times and have the benefit of having all possible amplitude values available to create codewords for the entire dataset.



\begin{algorithm}
    \caption{Two Pass Vector Quantization}\label{alg:2pass}
    \resizebox{0.76\linewidth}{!}{%
        \begin{minipage}{\linewidth}
            \begin{algorithmic}
                \Require Set of amplitudes, $Q = \{\psi_0, \psi_1, \ldots,\psi_d \}$, from an uncompressed simulation of the quantum circuit.
                \State Initialize data array $K$
                \For{$\psi$ in $Q$}
                    \For{amplitude, $A$, in $\psi$}
                        \State Append $A$ to $K$
                    \EndFor
                \EndFor
                \State Call k-means on data in K. Return Centroids (codewords), $C$
                \For{t from 0 to d-1}
                    \State Apply gate, $G_{t+1}$, to $\psi_t$: $\psi_{t+1}=G_{t+1} \cdot \psi_t$ using Algorithm \ref{alg:Gate}
                    \For{amplitude, $A$ in $\psi_{t+1}$}
                        \State Update $A$ to its nearest codeword: $A \leftarrow \arg\min_j \left\|A-C_j\right\|^2$
                    \EndFor
                \EndFor
            \end{algorithmic}
        \end{minipage}%
    }
\end{algorithm}


\Section{Experimental Results and Discussion}

\par To evaluate the effects of quantization on quantum simulations, we simulated repeated applications of the Quantum Fourier Transform (QFT). The QFT circuit is an appealing benchmarking candidate because it is a common sub-routine of many important, widely-used quantum algorithms such as Shor’s factoring algorithm, quantum phase estimation, and the hidden subgroup problem \cite{jordan2022quantum}. The QFT only requires Hadamard and controled-rotation gates, and it is a popular benchmark circuit \cite{wu2019compression,zulehner2019advanced,betelu2020limits}. In this paper, we simulate the QFT repeated 31 times on $6$ qubits for a total circuit depth of $d=651$. Our initial state vector, $\ket{\psi_0}$, is a superposition of computational basis states. We assess our lossy simulations by evaluating the fidelity between the simulated state vector and an \say{ideal} state vector computed analytically, then stored in complex double precision (128 bits/complex number). Lower precision floats are upcast to complex double precision for the fidelity calculation.


\SubSection{Scalar Quantization}

\par
\begin{figure}
\begin{center}
\begin{tabular}{cc}
\epsfig{height=2in,file=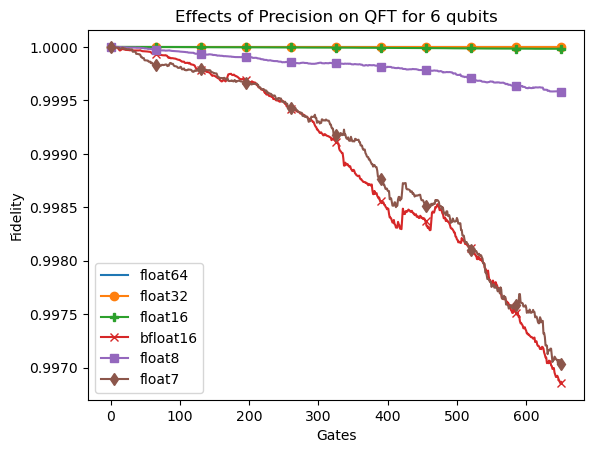} &
\epsfig{height=2in,file=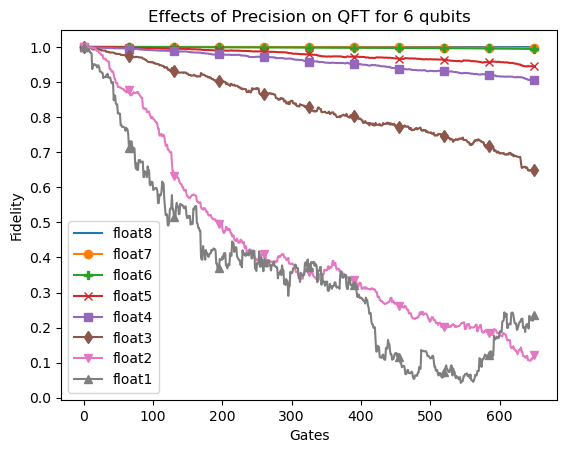} \\
{\small (a) float64 - \say{float7}} & {\small (b) \say{float8} - \say{float1}}
\end{tabular}
\end{center}
\caption{\label{fig:scalar}%
Scalar quantization simulation results.}
\end{figure}

 Scalar quantization works well in simulating deep quantum circuits. Figure \ref{fig:scalar} shows the results of our simulations. Figure \ref{fig:scalar} (a) displays the \say{high precision} results. For all precisions above \say{float7}, fidelities remained well above $0.99$ for the duration of the circuit. We see little distinction between double, single, and half precision truncated simulation. Interestingly, bfloat16 noticeably under-performs not only float16, which has access to the same number of bits, but also \say{float8}, a 16-bit float restricted to only $8$ of its significand bits. bfloat16's performance is closer to that of \say{float7}; while they share the same number of significand bits, bfloat16 has access to more exponent bits (Figure \ref{fig:precision}). This indicates that significand bits are more important than exponent bits when simulating quantum data. This is likely due to the irregular distribution of floating point numbers, which jump by a factor of $2$ when the exponent increases \cite{betelu2020limits}. Figure \ref{fig:scalar} (b) displays reducing the precision of the significand below $8$ bits, all the way down to a single bit. We can see that precisions \say{float4} and above maintain simulation fidelities $>0.9$ after hundreds of gate applications; however, below \say{float4} the simulation fidelity rapidly declines.

 Still, these results are promising, as such high fidelities at $4$ bits of precision in the significand indicates that lossy simulations of quantum circuits can perform well while using few bits. The relative insignificance of the exponent bits when simulating quantum circuits suggest that a non-traditional data format based on, for example, algebraic representation of complex numbers \cite{tsai2016bit} or their complex logarithm \cite{betelu2020limits} may be a more efficient use of bits than the floating point format. Furthermore, the success of extreme significand-constrained floats in this work indicates that such custom data formats may be able to use less bits than previously thought.

 \SubSection{Vector Quantization}

\par
\begin{figure}
\begin{center}
\begin{tabular}{cc}
\epsfig{height=2in,file=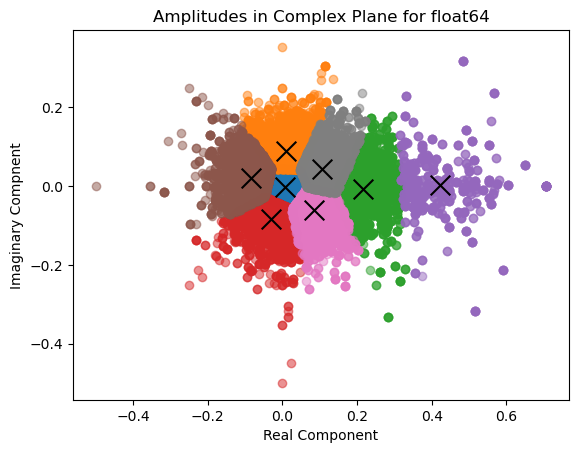} &
\epsfig{height=2in,file=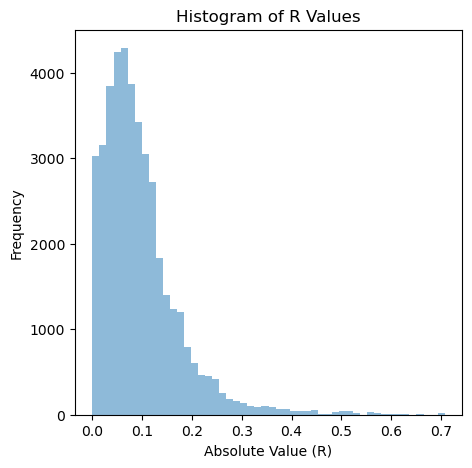} \\
{\small (a) Distribution of complex amplitudes} & {\small (b) Histogram of amplitude magnitudes}
\end{tabular}
\end{center}
\caption{\label{fig:dist}%
Distribution of simulation amplitudes. (a) shows every complex amplitude of every state vector in the simulation for double precision ($2^{n}(d+1)$ amplitudes) and the codewords, $X$s, selected by k-means and their associated clusters. $m=3$ above for a total of $2^3=8$ codewords. (b) is a histogram of the magnitudes, $R$, of the numbers in (a).}
\end{figure}

 We implement vector quantization by accounting for all possible complex amplitudes in the simulation (Figure \ref{fig:dist} (a)), then assigning codewords based on k-means as outlined in Algorithm \ref{alg:2pass}. The $2^m$ codewords themselves are stored in complex double precision format, but since a one-to-one mapping exists between the codewords and length $m$ bit strings, only $m$ bits are needed to represent a codeword. Since arithmetic operations (e.g. gate application) must be done at some precision, we employ vector quantization \textit{in addition} to scalar quantization. We first employ Algorithm \ref{alg:Gate} to evolve a state vector at the desired precision, then we round each amplitude to its nearest codeword in the complex plane. That result is in turn used as the input to the next state vector evolution. At any given time $t$, the state vector $\ket{\psi_t}$ can be represented using $m2^{n}$ bits where $2^m$ is the number of codewords. 

\par
\begin{center}
\begin{figure}
\centering
\begin{tabular}{ccc}
\centering
\epsfig{height=1.5in,file=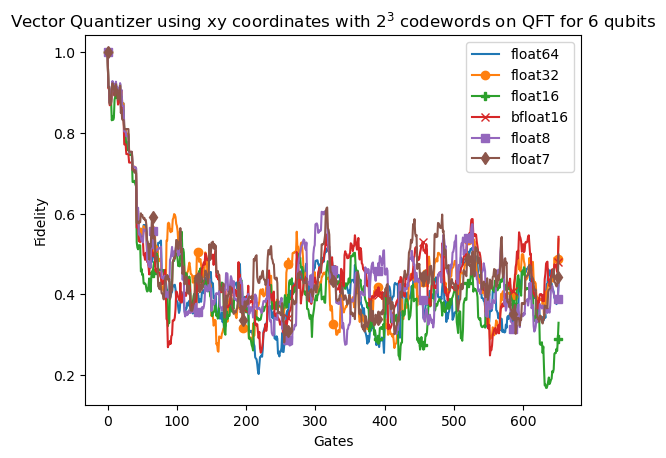} &
\epsfig{height=1.5in,file=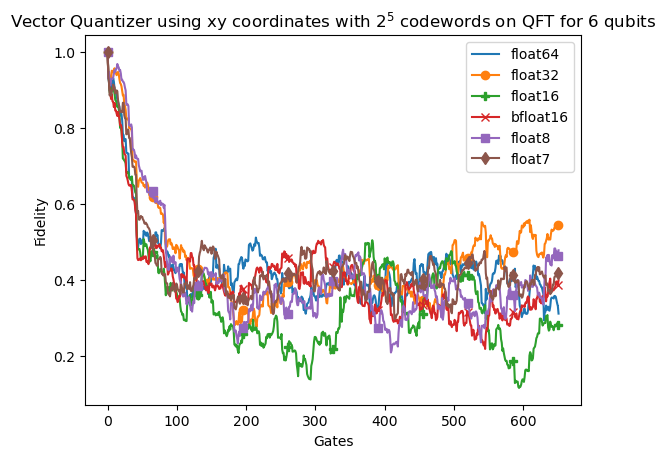} & 
\epsfig{height=1.5in,file=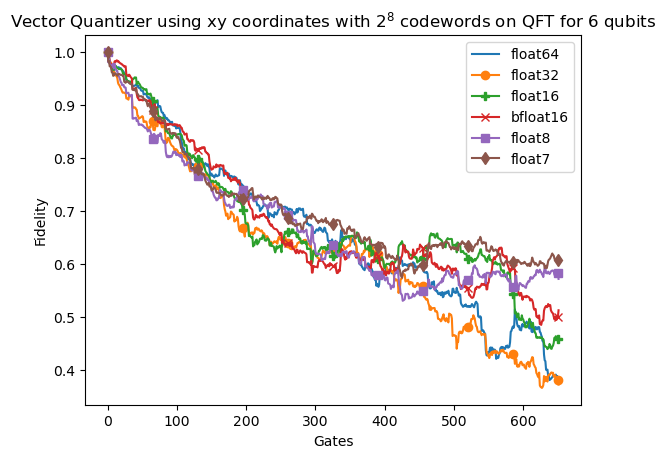} \\
\epsfig{height=1.5in,file=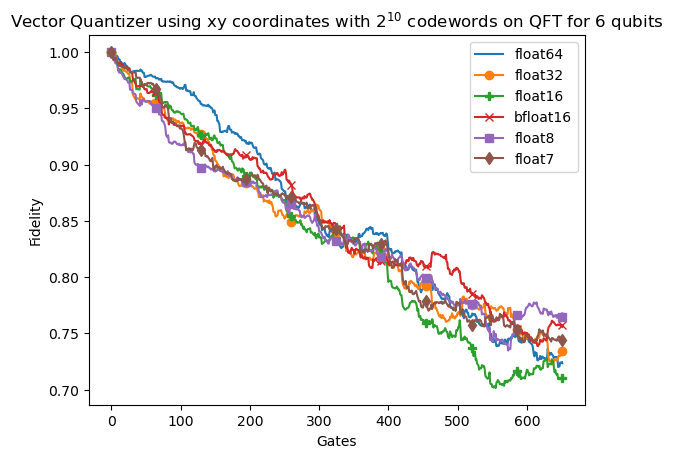} &
\epsfig{height=1.5in,file=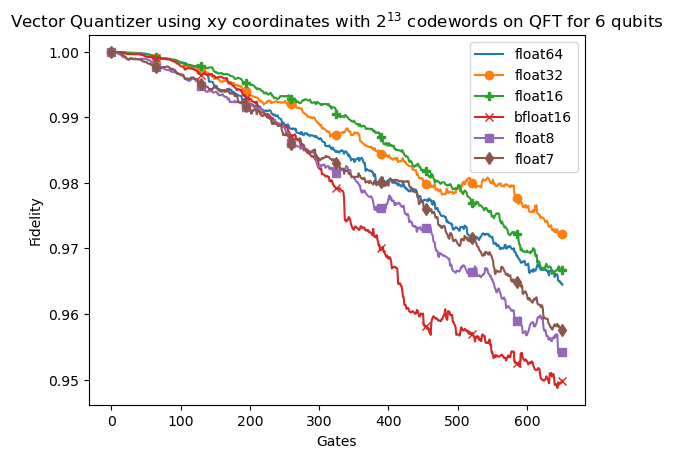} & 
\epsfig{height=1.5in,file=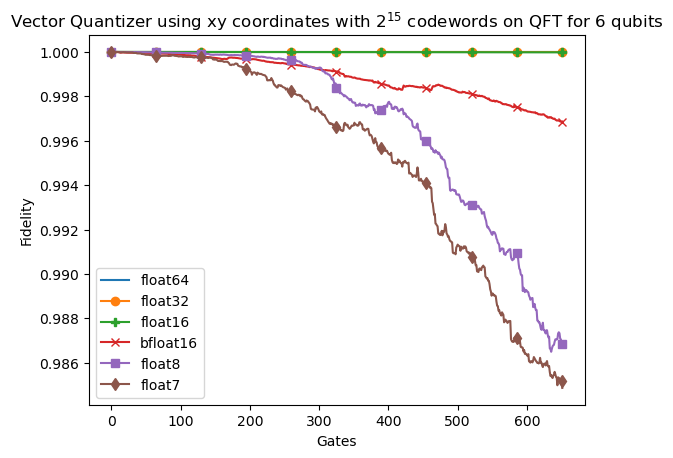} \\
\end{tabular}
\caption{\label{fig:codewords_fidelity}%
Computed fidelities for $m=[3,5,8,10,13,15]$. As the number of codewords increases, the simulation fidelity improves, especially at large depths.}
\end{figure}
\end{center}

 We simulated the QFT for $m \in \left[2,15\right]$. Results for $m = [3,5,8,10,13,15]$ are shown in Figure \ref{fig:codewords_fidelity}. Simulation fidelities quickly decay when few codewords are used, but improve with the number of codewords. At $2^{13}$ codewords, fidelities for all of the tested precisions are $>0.95$ and at $2^{15}$ codewords, fidelities are all $>0.98$. While the number of codewords remains low, the underlying precision used for the arithmetic has less of an impact on the simulation fidelity, as the ultra-low number of codewords is the largest source of error. This is why all precisions demonstrate similarly noisy fidelity results at low codeword numbers. However, when the number of codewords increases, while some noise due to the random initialization of the k-means centroids is present, we begin to see the higher precision floats perform better in general, until at $2^{15}$ codewords, the results are ordered with the precision of the underlying arithmetic operations. 

\par
\begin{figure}
\begin{center}
\begin{tabular}{ccc}
\epsfig{width=2in,file=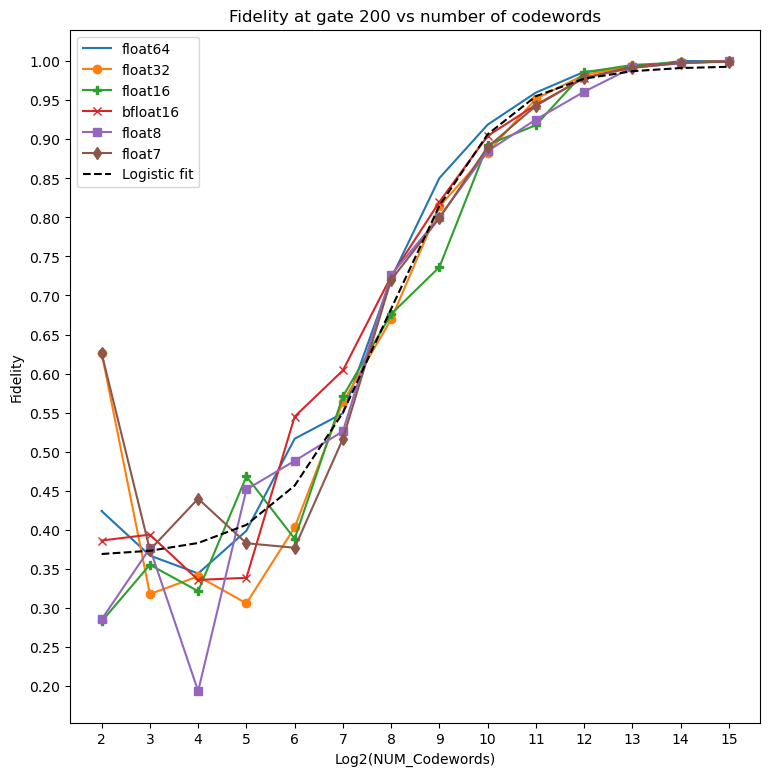} & 
\epsfig{width=2in,file=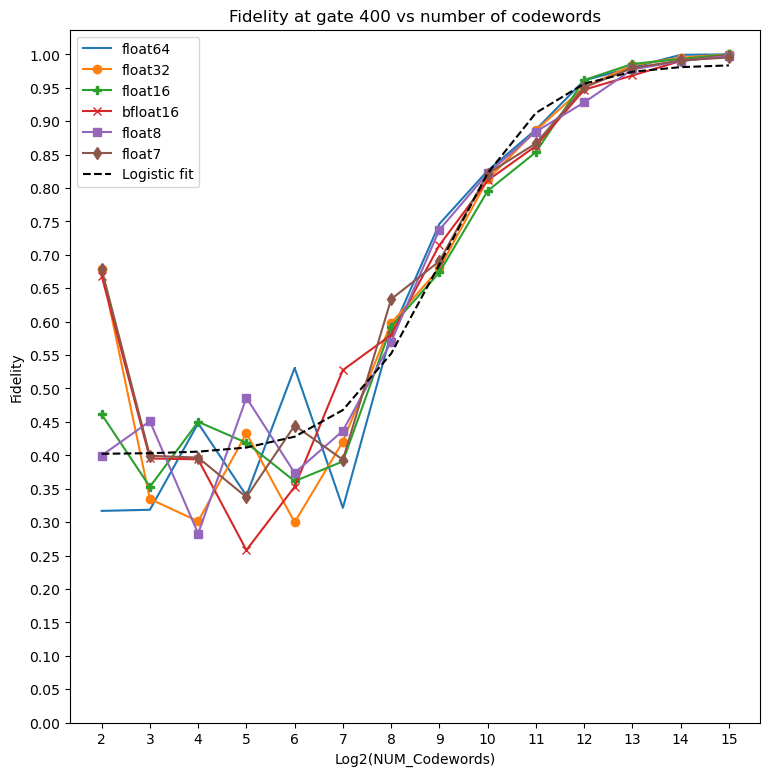} &
\epsfig{width=2in,file=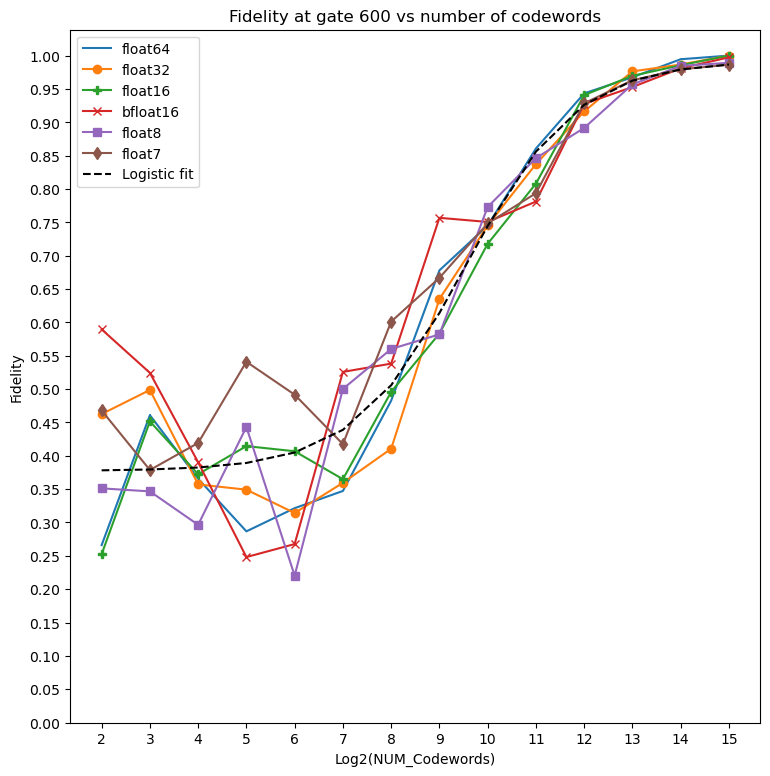} \\
\end{tabular}
\end{center}
\caption{\label{fig:logistic}%
Simulation fidelity as a function of the number of codewords used for vector quantization, shown (left to right) after gate 200, 400, and 600. The dashed line is a logistic fit of the mean of the results across precisions.}
\end{figure}

 We next investigate \textit{how} the fidelity changes with respect to the number of codewords. Figure \ref{fig:logistic} shows slices of the circuit simulation taken after sequential steps through the circuit of 200 gates. Just like in Figure \ref{fig:codewords_fidelity}, we can see that regardless of circuit depth, noise dominates at small number of codewords due to the error introduced by truncating to so few codewords, but as the number of codewords increases, the results become less noisy. From this relationship between the number of codewords used for vector quantization and the simulation fidelity, we can derive an upper-bound estimate of the number of codewords, $2^m$, needed to achieve a desired fidelity at a given depth. First, we empiricaly fit the following logistic function to the data in Figure \ref{fig:logistic} for each time step in the circuit simulation:

\begin{equation}
\text{fidelity}(m)=\frac{A}{1+e^{-k\left(m-x_0\right)}}+\text{off}
\label{eq:logistic_fit}
\end{equation}

\begin{figure}
\centering
\begin{tabular}{cccc}
\epsfig{width=1.5in,file=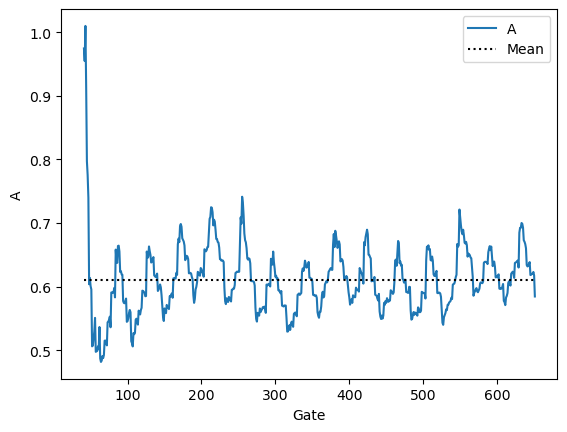} &
\epsfig{width=1.5in,file=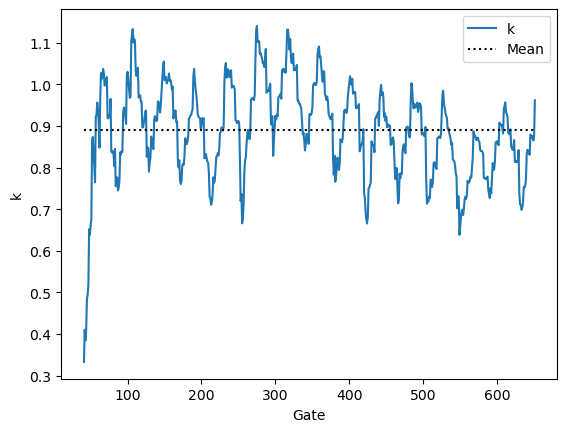} &
\epsfig{width=1.5in,file=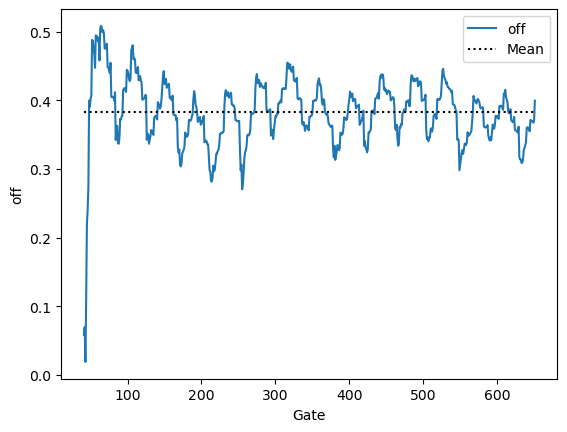} &
\epsfig{width=1.5in,file=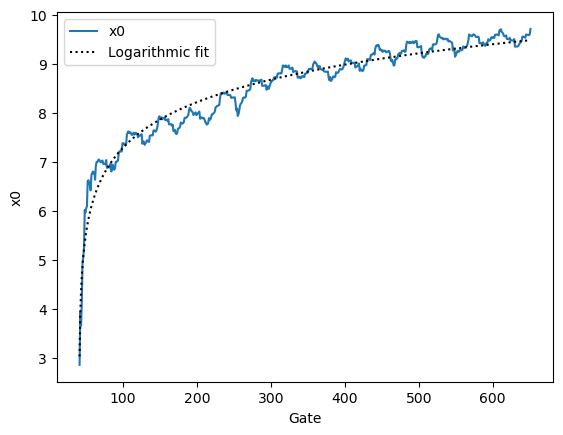} \\
{\small (a) A} & {\small (b) k} & {\small (c) off} & {\small (d) $x_0$}\\
\end{tabular}
\caption{\label{fig:fit_params}%
Parameters of the logistic fit $\text{fidelity}(m)=\frac{A}{1+e^{-k\left(m-x_0\right)}}+\text{off}$. $m=\log_{2}\text{\#codewords}$}
\end{figure}

\begin{figure}
\begin{center}
\begin{tabular}{cc}
\epsfig{height=2.2in,file=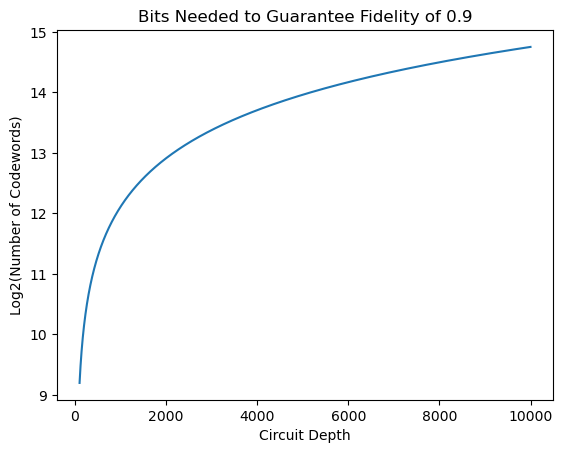} &
\epsfig{height=2.2in,file=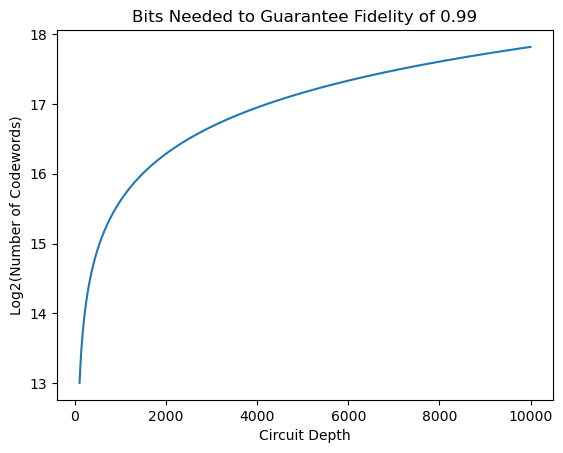} \\
{\small (b) $f=0.9$} & {\small (c) $f=0.99$}
\end{tabular}
\end{center}
\caption{\label{fig:fidelity_guarantee}%
Estimated number of codewords needed to achieve a given fidelity as a function of circuit depth, plotted for $d\in[10^{2},10^{4}]$. Using a vector quantization scheme no more than $2^{15}$ codewords should be needed to achieve simulation fidelities $\approx0.9$ at $10^4$ depth, and no more than $2^{18}$ codewords should be needed to achieve simulation fidelities $\approx0.99$.}
\end{figure}

Figure \ref{fig:fit_params} shows how the parameters of this fit change with the depth of the circuit being simulated. For three out of the four parameters, their asymptotic behavior with respect to the circuit depth is constant with some high-frequency noise. However, $x_0$ grows logarithmically with the circuit depth while sharing the same high-frequency noise as the other three parameters. We can invert equation (\ref{eq:logistic_fit}) to model the number of codewords as a function of the fidelity, $A$, $k$, and, the offset, off. If we model $A$, $k$, and off as constants, and $x_0$ as logarithmically increasing with circuit depth, then, for a desired fidelity, $f$, we can model the asymptotic behavior of the number of codewords needed to simulate $f$ as a function of circuit depth. Figure \ref{fig:fidelity_guarantee} shows the results of such a model. We can estimate the number of codewords needed in our vector quantization scheme to simulate a circuit of some depth, $d$, to a desired fidelity. $m$ asympotically increases as $O(\log d)$. Since a state vector can be represented in $m2^{n}$ bits, this means that to simulate a circuit while maintaining a given fidelity, the size of the stored state vectors should only scale logarithmically with the depth of the circuit to be simulated.

\Section{Conclusion}

\par In this work we study the lossy Schrödinger-style simulation of quantum circuits by applying both scalar and vector compression to reduce the storage overhead required for these kinds of simulations. We succeed in simulating $d=651$ depth Quantum Fourier Transform circuits on $n=6$ qubits to fidelities $>0.99$ by scalar quantizing the real and imaginary components of the complex amplitudes of state vectors. Further,  we maintain simulation fidelities $>0.9$ while truncating the significand of a float to only 4 bits. We also show that vector quantization can maintain fidelities $>0.98$ with $2^{15}$ codewords. We also utilize the way that the simulation fidelities improve with additional codewords to predict the behavior of the number of codewords necessary to achieve a desired simulation fidelity for a circuit of given depth. 

\par Our work leaves open several opportunities for future investigation. First, the results of our scalar quantized simulations indicate that exponent bits in floating point numbers may be less relevant to simulating quantum data, and the development of a quantum domain-specific data format and hardware build to handle such a format natively to could prove invaluable to quantum simulations, especially in the NISQ era, where we can not rely on large, fault-tolerent quantum devices. Second, in this work, we implement a \say{two-pass} vector quantizer, which first runs the circuit without compression, then uses that distribution of amplitudes to assign codewords. This is useful for circuits which are run frequently, but for infrequent circuits, a \say{one-pass} solution may be called for wherein codewords are assigned after every gate application while the circuit is running. This has $O(d)$ more calls to the k-means codeword finding subroutine, but avoids the need to run the circuit once without vector quantization. Third, our lossy quantization methods can be applied to other quantum simulation methods such as Feynman path integral simulation or tensor networks. Finally, more work can be done to compress the quantized state vectors post-simulation via lossless compression such as gzip, or state-of-the-art lossy compressor like SZ \cite{wu2019compression}.

\newpage

\Section{Acknowledgements}
\par The authors would like to thank Shubham Chandak, Katharine Hyatt, Qingxi Meng, Dorsa Fathollahi, and Pulkit Tandon for helpful conversations.

\par Some of the computing for this project was performed on the Sherlock cluster. We would like to thank Stanford University and the Stanford Research Computing Center for providing computational resources and support that contributed to these research results.
\par This research was supported by the National Science Foundation under Grant No. 2106508. Any opinions, findings, and conclusions or recommendations expressed in this material are those of the authors and do not necessarily reflect the views of the National Science Foundation.

\Section{References}
\bibliographystyle{IEEEbib}
\bibliography{refs}

\end{document}